\documentclass{jfm}
\usepackage{graphicx}
\usepackage{epstopdf, epsfig}
\usepackage{eurosym}
\usepackage{graphicx}
\usepackage{caption,subcaption}
\usepackage{float}
\usepackage[normalem]{ulem}
\usepackage{color}
\usepackage{soul}

\usepackage{bbold}
\usepackage{amsmath,amssymb}
\usepackage{braket}
\usepackage[english]{babel}
\usepackage{ulem}
\usepackage{cancel}

\newcommand{\beq}{\begin{equation}}
\newcommand{\eeq}{\end{equation}}
\def\mP{ \mathcal{P} }
\def\mD{ \mathcal{D} }
\def\mV{ \mathcal{V} }
\def\mE{ \mathcal{E} }
\def\mH{ \mathcal{H} }

\def\mI{ \mathcal{I} }
\def\mZ{ \mathcal{Z} }
\def\mT{ \mathcal{T} }
\def\mA{ \mathcal{A} }

\def\kmax{k_{max}}
\def\kt{k_{tran}}

\def\la{ \left\langle }
\def\ra{ \right\rangle }

\newcommand{\tu}{\tilde{ u}}

\newcommand{\bu}{{\bf u}}

\newcommand{\bk}{{\bf k}}
\newcommand{\bq}{{\bf q}}
\newcommand{\bp}{{\bf p}}

 \def\NEW#1{{\textcolor{black}{#1}}}

\shorttitle{Energy fluxes in quasi-equilibrium flows}
\shortauthor{A. Alexakis and M-E. Brachet}

\title{Energy fluxes in quasi-equilibrium flows}

\author{Alexandros Alexakis \aff{1}
  \corresp{\email{alexakis@lps.ens.fr}},
Marc-Etienne Brachet \aff{1}}

\affiliation{ \aff{1} Laboratoire de Physique de l'\'Ecole Normale Sup\'erieure, 
    ENS, Universit\'e PSL, CNRS, Sorbonne Universit\'e, Universit\'e de Paris, 
    F-75005 Paris, France }

\begin{document}

\maketitle
\begin{abstract}
We examine the relation between the absolute equilibrium state of the spectrally truncated Euler equations (TEE) predicted by Kraichnan (1973) to the forced and dissipated flows of the spectrally truncated Navier-Stokes (TNS) equations. 
\NEW{ In both of these idealized systems a finite number of Fourier modes is kept contained inside a sphere of radius $\kmax$ but while the first conserves energy in the second energy is injected by a body-force $\bf{f}$ and dissipated by the viscosity $\nu$.
For the TNS system stochastically forced with energy injection rate $\mI_\mE$    
we show, using an asymptotic expansion of the Fokker-Planck equation, that in the limit of small 
$\kmax\eta$ (where $\eta=(\nu^3/\mI_\mE)^{1/4}$ the Kolmogorov lengthscale) 
the flow approaches the absolute equilibrium solution of Kraichnan with such  an effective  ``{\it temperature}" so that there is a balance between the energy injection and the energy dissipation rate. }
We further investigate the TNS system using direct numerical simulations in periodic cubic boxes of size $2\pi/k_0$. 
\NEW{
The simulations verify the predictions of the model for small values of $\kmax\eta$.
For intermediate values of $\kmax\eta$ a transition from the quasi-equilibrium ``{\it thermal}" state to Kolmogorov turbulence is observed.
In particular we demonstrate that, at steady state, the TNS reproduce the Kolmogorov energy spectrum if $\kmax \eta \gg 1$. 
As $\kmax\eta$ becomes smaller then a bottleneck effect appears taking the form of the equipartition spectrum $E(k) \propto k^2$ at small scales. As $\kmax\eta$ is decreased even further so that $\kmax\eta \ll (k_0/\kmax)^{11/4} $ the equipartition spectrum occupies all scales approaching the asymptotic equilibrium solutions found before.} If the forcing is applied at small scales and the dissipation acts only at large scales then the equipartition spectrum appears at all scales for all values of $\nu$. In both cases a finite forward or inverse flux is present even for the cases where the flow is close to the equilibrium state solutions. 
\NEW{ However, unlike the classical turbulence where an energy cascade develops with a mean energy flux that is large compared to its fluctuations,  
the quasi-equilibrium state has a mean flux of energy 
that is subdominant to the large flux fluctuations observed.
}


\end{abstract}


\section{Introduction}        
\label{sec:intro}             

Turbulence is a classical example of an out of equilibrium system. In steady state, energy is constantly injected at some scale $\ell_{in}$ while it is dissipated at smaller scales $\ell_{\nu}$ by viscous forces. This process requires a finite flux of energy from the former scale $\ell_{in}$ to the latter $\ell_{\nu}$ that is provided by the well known Kolmogorov-Richardson cascade.
Despite the out of equilibrium nature of turbulence there are circumstances where equilibrium dynamics become relevant. 
This has been claimed to be the case for the scales larger than than the injection scale $\ell> \ell_{in}$.  At these scales the energy flux is zero and can possibly be modeled using equilibrium dynamics \citep{dallas2015statistical, cameron2017effect, alexakis2018thermal}.
\NEW{Furthermore,  
at the smallest scales of the inertial range a so-called `{\it bottleneck}' 
manifests where the power-law  slope of the energy spectrum becomes less steep
\citep{donzis2010bottleneck, falkovich1994bottleneck, martinez1997energy, lohse1995bottleneck}. This has been interpreted by \cite{frisch2008hyperviscosity}
as an `incomplete thermalization' that becomes asymptotically at equilibrium 
for hyper-viscous flows when the order of the hyper-viscosity tends to infinity.
}
Equilibrium dynamics become also relevant  in the presence of inverse cascades in finite domains where large scale condensates form  \citep{kraichnan1967inertial, robert1991statistical, naso2010statistical, bouchet2012statistical, shukla2016statistical}. Finally understanding equilibrium dynamics is important for systems that display a transition of from forward to an inverse cascade 
\citep{alexakis2018cascades, 
     sahoo2017discontinuous, 
      benavides2017critical,
      sozza2015dimensional,
     deusebio2014dimensional, 
     seshasayanan2014edge} 
since the large scale flows in these systems transition from an equilibrium state to an out of equilibrium state.
Besides the possible applications, understanding the equilibrium dynamics in turbulence is also a much needed step required before  understanding its much harder out-of-equilibrium counterpart. 
This has led many researchers \citep{LEE:1952p4100, hopf1952statistical, kraichnan1967inertial, kraichnan1973helical, OrszagHouches} to investigate the equilibrium state of the truncated Euler equations (TEE) where only a finite number of Fourier modes is kept and are given by:
\beq
\partial_t \bu + \mathbb{P}_K[\bu \cdot \nabla \bu +\nabla p] =0,
\qquad \nabla\cdot \bf u=0.
\label{TEE}
\eeq
Here $\bf u$ is the incompressible velocity field, $p$ is the pressure and $\mathbb{P}_K$ is a projection operator 
that sets to zero all Fourier modes except those that belong to a particular set `$K$' (here chosen to be all 
wavenumbers inside a sphere centered at the origin with radius $k_{max}$). 
These equations conserve exactly two quadratic invariants 
\beq 
\mathrm{the\,\, Energy} \quad \mathcal{E}=\frac{1}{2}\int |{\bf u}|^2 dx^3 \quad \mathrm{and\,\,the \,\, Helicity} \quad \mathcal{H}=\frac{1}{2}\int {\bf u\cdot\nabla \times u} dx^3.
\label{energy} 
\eeq
In Fourier space these invariants are distributed among the different modes which are quantified by the energy and helicity spherically averaged spectra $E(k),H(k)$ respectively defined as
\beq \NEW{
E(k)=\frac{1}{2k_0}\sum_{k\le |{\bf k}| <k+k_0} |\tilde{\bf u}_{\bf  k}|^2 \quad \mathrm{and} \quad  
   H(k)=\frac{1}{2k_0}\sum_{k\le |{\bf k}| <k+k_0}  \tilde{\bf u}_{\bf -k}\cdot(i{\bf k} \times \tilde{\bf u}_{\bf k}).  }
\eeq
\NEW{Here $\tilde{\bf u}$ is the Fourier transform of $\bf u$ and we have assumed a triple periodic cubic domain of size $2\pi/k_0$. The spectra have been divided by the smallest non-zero wavenumber $k_0$
so that they have units of energy and helicity density respectively.}

At late times this system reaches a statistically steady state whose properties are fully determined by these two invariants. Using Liouville's theorem and assuming ergodicity \cite{LEE:1952p4100} predicted that at {\it absolute} equilibrium this system will be such that every state $\bu$ of a given energy $\mE$ is equally probable. This is equivalent to the micro-canonical ensemble in statistical physics and it leads to equipartition of energy among all the degrees of freedom (ie among all Fourier amplitudes) and an energy spectrum given by $E(k)\propto k^2$.
\cite{kraichnan1973helical} generalized these results including helicity and assuming a Gaussian equipartition ensemble 
\beq
\mP(\bu) = \mZ^{-1} \exp\left[ - \alpha \mE -\beta \mH \right]  
\label{KR}
\eeq
where $\mP(\bu)$ is the probability distribution for the system to be found in the state $\bu$, $\mE$ is the energy and $\mH$ the helicity given in \ref{energy} and $\mZ$ a normalization constant.
The parameters $\alpha$  and $\beta$ are the equivalent of an inverse temperature and inverse chemical potential
in analogy with statistical physics. 
\NEW{We note that for the TEE system eq.\ref{KR} is not exact! This is because eq. \ref{KR} allows for fluctuations of the energy $\mE$
which are not allowed for the TEE system. However it becomes closer to the true distribution as the number N of fourier modes
becomes larger.}
For no helicity $\beta=0$ this leads to the energy spectrum $E(k)\propto k^2$ predicted by Lee.  In the presence of helicity  one obtains
\beq
E(k) = \frac{4\pi \alpha k^2}{\alpha^2 - \beta^2 k^2 },\qquad  H(k) = \frac{4 \pi \beta k^4}{\alpha^2 - \beta^2 k^2 }.
\label{eq:EkTh}
\eeq 
The  coefficients $\alpha$ and $\beta$ are determined by imposing the conditions
\beq
\NEW{\mathcal{E}=k_0 \sum_{\bf k} E(k) \quad \mathrm{and}\quad  \mathcal{H}= k_0 \sum_{\bf k} H(k) }
\eeq
and $\mathcal{E}$ and $\mathcal{H}$ are the initial energy and helicity respectively.    
The predictions above have been verified for the truncated Euler system in numerous numerical simulations \citep{OrszagPatterson72,cichowlas2005effective, krstulovic2009cascades, dallas2015statistical, cameron2017effect, alexakis2018thermal}. 

In the TEE however there is no exchange of energy with external sources or sinks and \def\mT{ \mathcal{T} }it is thus harder to make contact with the more realistic systems mentioned in the beginning of the introduction such as the scales larger than the forcing scale in a turbulent flow.  
It was shown recently in \cite{alexakis2018thermal} that, in some cases, although the large scales are close to an equilibrium state there is still exchange of energy with the smaller turbulent and forcing scales generating energy fluxes (from the forced scales to the large scales and from the large scales to the turbulent scales). 
It appears thus that, even in the presence of sources and sinks, equilibrium dynamics can still be relevant.
In this work we examine further this possibility by looking at the truncated Navier-Stokes (TNS) equations where there is constant energy injection and dissipation like in the regular Navier-Stokes Equation but the system is limited to a finite number of Fourier modes as in TEE. 
%
%
\NEW{
We show analytically in the next section that for weak energy injection and weak viscosity
so the $\kmax\eta \ll (k_0/\kmax)^{11/4}$ (where $\eta$ the kolmogorov lengthscale)
the system indeed reaches a quasi-equilibrium state whose probability distribution $\mP(\bu)$ can be calculated.} We verify and extend these results using direct numerical simulations in section \ref{sec:DNS}. Our conclusions are presented in the last section.


\section{Asymptotic expansion     }        
\label{sec:asmpt}             
 
We consider the truncated  Navier-Stokes (TNS) equations
\beq
\partial_t \bu + \mathbb{P}_K[\bu \cdot \nabla \bu +\nabla P] = {\bf f} + \nu \Delta \bu \label{TNS}.
\eeq
where $\bu$ is the incompressible velocity field, $P$ is the pressure, $\nu$ is the viscosity
and $\bf f$ is a forcing function. \NEW{The domain is a $2\pi$ periodic cube so that the smallest non-zero wavenumber is $k_0=1$.} The projection operator $\mathbb{P}_K$ sets to zero all Fourier modes with wavenumbers 
outside  the sphere of radius $\kmax$. \NEW{ In total there are $N\simeq \frac{4\pi}{3}(k_{max}/k_0)^3$ Fourier wavenumbers inside this sphere.} In order to proceed it helps to write the truncated Navier Stokes equation in Fourier space using the Craya-Lesieur-Herring decomposition (\cite{craya1958contributiona, Lesieur72, herring1974approach}) where every Fourier mode is written as the sum of two modes one with positive helicity and one with negative helicity 
$ \tilde{\bf u}_\bk = \tilde{u}_\bk^+ {\bf h}^+_\bk +   \tilde{u}_\bk^- {\bf h}_\bk^- $.
The two vectors $\bf h_k^\pm$ are given by: 
\beq
{\bf h}^s_{\bf k}= \frac{\bf k\times (\hat{e} \times k) }{\sqrt{2}\bf|k \times (\hat{e} \times k)|}
  +  i \, s\frac{\bf \hat{e} \times k }{\sqrt{2}\bf |\hat{e} \times k|} 
\eeq
where $\bf \hat{e}$ is an arbitrary unit vector. 
The sign index $s=\pm1$ indicates the sign of the helicity of ${\bf h}^s_{\bf k}$.
The basis vectors ${\bf h}^s_{\bf k}$ are eigenfunctions of the curl operator in Fourier space such that $i{\bf k \times h}^s_{\bf k} = s |{\bf k}| {\bf h}^s_{\bf k}$.
They satisfy   ${\bf h}^s_{\bf k} \cdot {\bf h}^s_{\bf k} =0$ and  $({\bf h}^s_{\bf k})^* \cdot {\bf h}^{s}_{\bf k}=1$, where the complex conjugate of ${\bf h}^s_{\bf k}$ is given by $({\bf h}^s_{\bf k})^*={\bf h}^{-s}_{\bf k}={\bf h}^s_{-\bf k}$. They form a complete base for incompressible zero-mean vector fields. This decomposition has been extensively used and discussed in the literature 
\citep{cambon1989spectral, 
       waleffe1992nature,
       chen2003joint, 
       biferale2012inverse, 
       moffatt2014note, 
       alexakis2017helically, 
       sahoo2017discontinuous}. 
Note that since every $\bf \tu_\bk$ is described by two complex 
amplitudes $\tilde{u}^\pm_\bk$ that satisfy $(\tilde{u}^\pm_\bk)^* = \tilde{u}^\pm_{-\bk}$ there are in total $2N$ independent degrees of freedom. The truncated Navier-Stokes using the helical decomposition can then be written as 
\beq
\partial_t \tu^{s}_\bk  = \mathcal{V}^s_\bk 
                          - \nu k^2   \tu^{s}_\bk  +  \tilde{ f}^s_\bk %
\eeq
where the nonlinear term $\mathcal{V}^s_\bk$ is written as the convolution 
\beq
\mathcal{V}^s_\bk = \sum_{\bf p+q=k}  \sum_{s_p,s_q} C^{s,s_q,s_p}_{\bk,\bq,\bp} \tu^{s_p}_\bp \tu^{s_q}_\bq
\eeq
and the tensor $C^{s_k,s_q,s_p}_{\bk,\bq,\bp}$ is given by
$
C^{s_k,s_q,s_p}_{\bk,\bq,\bp} =   \frac{1}{2} (s_qq-s_pp) ({\bf h}_{-\bk}^{s_\bk} \cdot  {\bf h}_\bq^{s_\bq} \times {\bf h}_\bp^{s_\bp})  .
$
The nonlinearity $\mathcal{V}^s_\bk$ satisfies the following relations
\beq
\sum_{s,\bk}     \tu^s_{-\bk} \mV^s_\bk =0,   \qquad
\sum_{s,\bk} s \, k \, \tu^s_{-\bk} \mV^s_\bk =0 
\eeq
that correspond to the energy and helicity conservation respectively and
\beq 
\sum_{s,\bk} \frac{\partial}{\partial \tu^s_\bk}  \mV^s_\bk =0.
\eeq
This last relation indicates that phase space volume is conserved by the nonlinearity ({\it ie} it satisfies a Liouville condition). We will assume that the forcing is written as $\tilde{ f}^s_\bk = {\epsilon}^s_\bk \xi^s_\bk$ where $\xi^s_\bk$ are random complex amplitudes that are statistically independent, normally distributed and delta-correlated in time such that $\langle \xi^s_\bk(t) \xi^{s'}_{-\bq}(t') \rangle= 2 \delta_{s,s'} \, \delta_{\bq,\bk} \, \delta(t-t')$. 
With this choice each forcing mode injects  energy to the system on average at rate $\epsilon^s_\bk$. 
Then the Fokker-Plank equation for the probability density $\mP(\bu)$ in the 2N-dimensional space of all complex amplitudes $\tilde{u}^\pm_\bk$ is given by
\beq
\frac{\partial}{\partial t} \mP + \sum_{s,\bk} \frac{\partial}{\partial \tu^s_\bk } (\mV^s_\bk \mP ) = 
               \nu \sum_{s,\bk}  \frac{\partial}{\partial \tu^s_\bk } ( k^2 \tu^s_\bk \mP ) + 
               \sum_{s,\bk} \epsilon^s_\bk  \frac{\partial}{\partial \tu^s_\bk }\frac{\partial}{\partial \tu^s_{-\bk} } \mP,
               \label{FP}
\eeq 
where the sum is over all 2N modes $\tu^s_\bk$.
Multiplying by $\mE=\frac{1}{2} \sum_{s,\bq} |\tu^s_\bq|^2$, integrating over the phase-space volume and using integration by parts we obtain the energy balance equation
\beq
\frac{\partial}{\partial t} \la \mE \ra = - \la \mD_\mE \ra + \la \mI_\mE \ra \label{enba}
\eeq  
where $\mD_\mE=\nu \sum_{s,\bk}  k^2 \langle |\tu^s_\bk|^2 \rangle$ is the energy dissipation,  
$\mI_\mE= \sum_{s,\bk} \epsilon^s_\bk$ is the injection rate 
and the brackets stand for the average $\la f \ra \equiv 
\int f \mP dU $ where $dU$ stands for the phase space volume element $dU = \prod_{s,\bk} d\tu^s_\bk$.  

We are interested in the limit that the energy injection and dissipation rate are a small perturbation to the thermalized fluctuations. We thus set $\nu_k = \delta \nu_k'$, $\epsilon^s_\bk =\delta \epsilon'^s_\bk$ where $\delta\ll1$ is a small parameter. 
We then expand $\mP$ in power series of $\delta$ as
 $ \mP = \mP_0 + \delta \mP_1 + \dots  $ . 
We are going to also consider the long time limit so we can neglect the time derivative. 
To zeroth order we then have
\beq
\sum_{s,\bk} \frac{\partial}{\partial \tu^s_\bk } (\mV^s_\bk \mP_0 ) =0. \label{trajectories} 
\eeq 
The equation above implies that $\mP_0$ is constant along the trajectories in the phase space followed by solutions of the truncated Euler equations. These trajectories are expected to be chaotic for large $N$ and since this is such a high dimensional space we can also conjecture that these trajectories are space filling (ie ergodic) in the subspace constrained by the invariants of the system. In other words we assume that the trajectory will pass arbitrarily close to any point that has the same energy and helicity as the initial conditions. 
In this case $\mP_0$  
is determined by the energy and helicity of the system $\mP(\bu)=f(\mE,\mH)$. For the present work however we are going to neglect the second invariant the helicity and assume dependence only on the energy. We then write the solution of eq. (\ref{trajectories}) as
\beq 
\mP_0 = f(\mE) = f\left( \frac{1}{2}\sum_{s,\bk} |\tu^s_\bk|^2 \right). 
\label{zerothsolution}
\eeq
 
To next order we then get 
\beq
               \sum_{s,\bk} \frac{\partial}{\partial \tu^s_\bk } (\mV^s_\bk \mP_1 ) = 
               \nu' \sum_{s,\bk}\frac{\partial}{\partial \tu^s_\bk } ( k^2 \tu^s_\bk \mP_0 ) + 
               \sum_{s,\bk} \epsilon'^s_\bk  \frac{\partial}{\partial \tu^s_\bk }\frac{\partial}{\partial \tu^s_{-\bk} } \mP_0
\eeq 
Substituting eq. (\ref{zerothsolution}) and using the chain rule for the derivatives 
\beq \frac{\partial}{\partial \tu^s_\bk } f(\mE) = 
     \frac{\partial \mE}{\partial \tu^s_\bk } \frac{\partial f}{\partial \mE } =
     \left( \frac{\partial}{\partial \tu^s_\bk } \frac{1}{2} \sum_{s_s,\bq} \tu^{s_q}_{-\bq} \tu^{s_q}_{\bq} \right) \frac{\partial f}{\partial \mE } =
     \tu^{s}_{-\bk} \frac{\partial f}{\partial \mE } 
\eeq
we obtain for the function $f(\mE)$
\beq
 \sum_{s,\bk} \frac{\partial}{\partial \tu^s_\bk } (\mV^s_\bk \mP_1 ) = 
  \nu' \sum_{s,\bk} \left(f + |\tu^s_\bk|^2 \frac{\partial f}{\partial \mE } \right) k^2 + 
   \left( \left[\sum_{s,\bk} \epsilon'^s_\bk \right] \frac{\partial f}{\partial \mE }   
          +  \left[ \sum_{\bk}\epsilon'^s_\bk |\tu^s_\bk|^2 \right]  \frac{\partial^2 f}{\partial \mE^2 }
            \right).  \label{1stOrd}
\eeq 
To obtain a closed equation for $f(\mE)$ we average eq. (\ref{1stOrd}) over the volume $dU_\mE$
of all points in phase space of energy between $\mE$ and $\mE+d\mE$.  This consists
of a spherical shell in the 2N-dimensional phase space of radius $\sqrt{2\mE}$.
Averaging over this volume leads the sum in the left hand side to drop out because it is a divergence
and the trajectories determined by $\mV^s_\bk$ stay inside the shell.
The volume integrals of terms independent of $\tu^s_\bk$ are proportional to the shell volume $dU_\mE = S_{2N} (2\mE)^{N-1} d\mE$ where $S_{2N}$ is the surface of an unit radius 2N-dimensional sphere. Terms proportional to $|\tu^s_\bk|^2$ result due to symmetry $\int_{dU_\mE} |\tu^s_\bk|^2 dU = (2N)^{-1} \int_{dU_\mE} \sum_{s,\bk} |\tu^s_\bk|^2 dU = (2N)^{-1} S_{2N} (2\mE)^{N} $.
This leads to
\beq
                            \nu' \left(\sum_\bk |\bk|^2 \right) \left(   f + \frac{\mE}{N}  \frac{\partial f}{\partial \mE } \right) + 
                                    \left[\sum_{s,\bk} \epsilon'^s_{\bk} \right] 
                                    \left( \frac{\partial f}{\partial \mE }   +  \frac{\mE}{N}  \frac{\partial^2 f}{\partial \mE^2 } \right)=0.
\eeq 
If we set \NEW{ $K^2 = \sum_{s,\bk} |\bk|^2 \simeq 8\pi \kmax^5/(5k_0^3)$} 
and $\mI_\mE'=\left[ \sum_{s,\bk}\epsilon'^s_{\bk} \right]$ 
then by multiplying by $\mE^{N-1}$ the equation simplifies to
\beq
\frac{\partial }{\partial \mE } \left( \nu' K^2  \mE^N   f  + 
 \mI_\mE' \mE^N  \frac{\partial f}{\partial \mE }  \right) =0
\eeq
that has the bounded solution:
\beq
f(\mE) = \mZ^{-1}\exp\left(-\frac{\nu K^2 }{\mI_\mE  } \mE \right)  \label{final}
\eeq
where $\mZ$ is a normalization constant that imposes $\int \mP(\bu) d\bu =1$ 
\NEW{ and is given by:
\beq \mZ = S_{2N} 2^{N-1} \left(\frac{\mI_\mE}{\nu K^2}  \right)^N \Gamma(N) \eeq
with $\Gamma$ being the Gamma function. }
We have thus recovered the Kraichnan distribution of eq. (\ref{KR}) with $\beta=0$ and inverse temperature given by $\alpha = \nu K^2/\mI_\mE$. Note that $\alpha$ depends only on the ratio of $\nu'$ and $\mI'_\mE$ and thus is independent of $\delta$ and we have thus dropped the primes. 

\NEW{
It worth restating that $\mP(\bu) = f\left(\frac{1}{2}\sum_{s,\bk} |\tu^s_\bk|^2\right)$ given in eq.\ref{final} expresses the probability 
that the system finds itself in the particular state $\bu$ with energy $\mE=\frac{1}{2}\sum_{s,\bk} |\tu^s_\bk|^2$. If we would like to find the probability $P(\mE)$ of finding the system in any state $\bu$ of energy $\mE$ we need to average over all states $\bu$ that have energy $\mE$. This leads to the chi-distribution for the energy 
\beq 
P(\mE) = \frac{S_{2N} (2\mE)^{N-1} }{\mZ} \exp\left(-\frac{\nu K^2 }{\mI_\mE  } \mE \right) \label{pdfEnergy}.
\eeq 
For large $N$ the distribution $P(\mE)$ in \ref{pdfEnergy}  is highly peaked at the mean energy 
\beq 
\langle \mE \rangle=\frac{N\mI_\mE}{\nu K^2}.  \label{eq:meanE}
\eeq
As $N$ tends to infinity $P(\mE)$ becomes asymptotically a delta function centered at $\langle \mE \rangle $.}
\NEW{ The mean energy of any mode $\tu_\bk^s$ is given by $ \frac{1}{2}\langle |\tu_\bk^s|^2 \rangle = \langle \mE \rangle/2N $
Averaging over spherical shells then leads to the thermal equipartition spectrum 
\beq E(k) = \frac{4\pi \mI_\mE}{\nu K^2 k_0^3} k^2  \label{fspec} \eeq
Finally using eq. (\ref{final}) one can calculate the energy dissipation
\begin{eqnarray}
\langle \mD_\mE \rangle &=& \nu \mZ^{-1} \sum_{s,\bq} q^2 \int  |\tu^s_\bq|^2 
        \exp\left(-\frac{1}{2 } \alpha \sum_{s,\bk}  |\tu^s_\bk|^2 \right) dU \nonumber \\
     &=& \frac{\nu}{2 N \mZ}  \left( \sum_{s,\bq} q^2 \right) \int (2\mE) \exp\left(-\alpha \mE \right) dU 
     \nonumber \\
     &=& \frac{\nu K^2}{N} \langle \mE \rangle =\mI_\mE
\end{eqnarray}
and verify that the energy balance relation  in eq. (\ref{enba}) is satisfied.}
The results indicate therefore that for small viscosity the truncated system will converge to the absolute  equilibrium solutions of such ``temperature" $1/\alpha$ so that the viscous dissipation balances the energy injection rate!

\section{Numerical simulations} 
\label{sec:DNS}               

\NEW{ 
In this section we test the results of the previous section and extend our investigation beyond the asymptotic limit using direct numerical simulations of the TNS system of eq. (\ref{TNS}). The simulations were performed using the {\sc ghost} code \citep{mininni2011hybrid} that is a pseudospectral code with 2/3 de-alliasing and a second order Runge-Kutta. For all runs the energy injection rate was fixed to unity and the  integration times were sufficiently long so that steady states were reached. The forcing used is random and white in time as the one discussed in the previous section. It is limited to a spherical shell of wavenumbers 
satisfying $(k_{_F} \le |\bk| \le k_{_F}')$. }

\NEW{
Three cases were examined.
In the first case small resolution runs were performed  on a cubic domain with $N_G=32$ grid points in each direction. 
These runs due to their small size allow to some extend a direct investigation of probability distribution function $\mP(\bu)$.}
In the second case  the simulations were performed on a larger grid $N_G=256$ numerical grid that, 
after de-aliasing, leads to $k_{max}=85$ and was forced at large scales $(k_{_F}=1,  k_{_F}'=2)$.
These simulations demonstrate the transition from a forward 
cascade to a quasi-equilibrium state predicted in the section before.
\NEW{
In the third case the simulations were designed to demonstrate the presence of an inverse flux in the thermalized state.
A smaller grid was used $N_G=128$ with $k_{max}=42$.  The energy injection was at large wavenumbers $(k_{_F}=31,  k_{_F}'=35)$ while we replaced $\nu \nabla^2 \bu$ by the modified viscous term $\nu \nabla^2 \mathbb{P}_{Q_{_D}} [\bu]$ that acts only a particular spherical set of small wavenumbers $Q_{_D}$ satisfying $(1 \le |\bk| \le 4)$.
With this setup energy is forced to be transported inversely from the forced wavenumbers to the dissipation wave numbers.  }

\subsection{ Small resolution runs }

\begin{figure}
\centering
\includegraphics[width=0.48\textwidth]{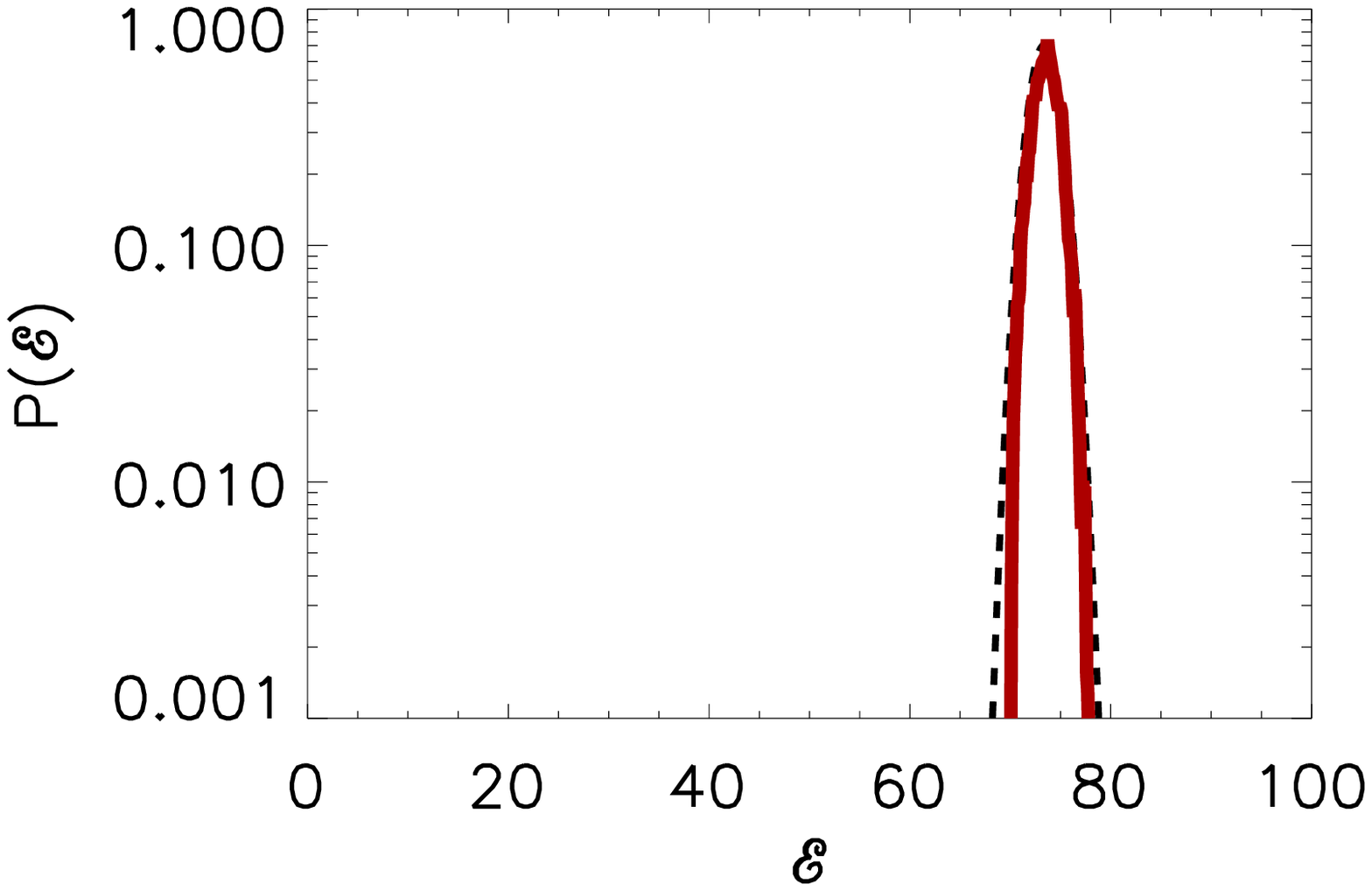}
\includegraphics[width=0.48\textwidth]{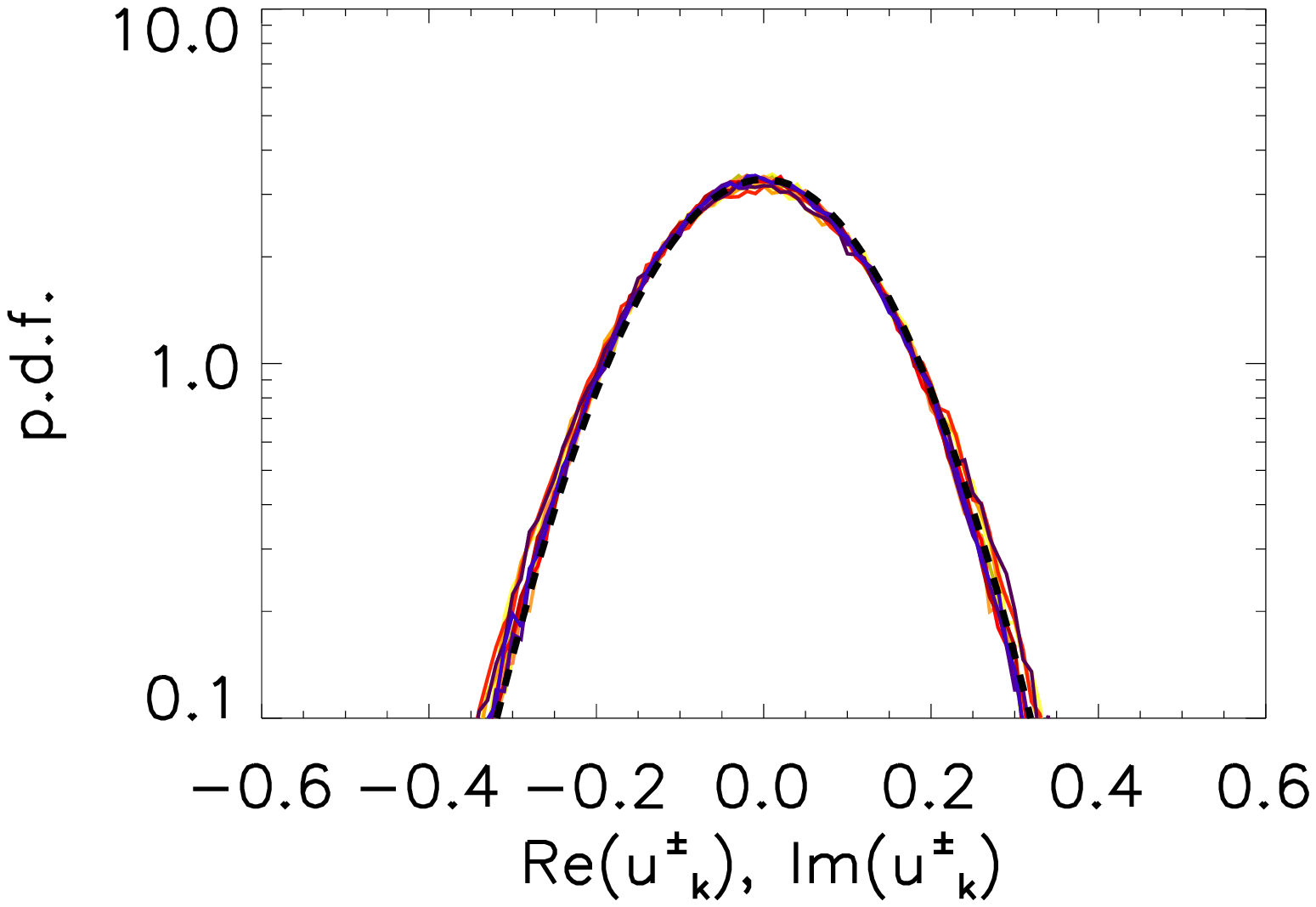}
\caption{Left panel: The probability distribution functions $P(\mE)$ from the results of the $32^3$-grid numerical simulations
compared with the theoretical prediction. Right panel: The probability distribution functions for the amplitudes of $\bu_\bk^s$ (real and imaginary part) for $\bk=(0,2,0),(0,4,0),(0,8,0)$, and for  $s=\pm 1$. }
\label{fig:pdf1}
\end{figure}

\NEW{
For these small resolution runs the energy injection rate was fixed to unity and the viscosity was set to $\nu=10^{-4}$.
The maximum wavenumber for the grid $N_G=32$ that  was used was $\kmax=10$. Despite the small value of $\kmax$ the total number of Fourier modes in this system is still quite high $N=5040$. It is thus still impossible to verify the predictions 
for $\mP(\bu)$ in full detain for such a high dimensional space.  
Nonetheless in the left panel of figure \ref{fig:pdf1} we plot the 
probability distribution function for the energy $P(\mE)$ based 
on the direct numerical simulations plotted with solid brown line
and based on the analytic predictions of \ref{pdfEnergy}.
The two curves overlap indicating that both the mean value 
and fluctuations around it are correctly captured.
Note that for this large value of $N$  the distribution is highly peaked 
however there is still a finite variance of $\mE$ indicating that $\mE$ 
is a fluctuating quantity at difference with the TEE where $\mE$ is fixed by the initial
conditions.}

\NEW{ To further explore the validity of the result \ref{final}
on the right panel of figure \ref{fig:pdf1} we plot the pdfs of the real and imaginary part
of the modes $\tu^s_\bk$ for the three wavenumbers $\bk=(0,2,0),(0,4,0),(0,8,0)$, and for  $s=\pm 1$.
The results of the previous section predict that these amplitudes follow a Gaussian distribution
with the same variance. There are 16 curves in total in the left panel of \ref{fig:pdf1} that have the same variance and 
perfectly overlap with the Gaussian shown by the dashed line in agreement the prediction
of eq. \ref{final}. Further more we calculated the
elements of the co-variance matrix for these modes $\Sigma{i,j} = \langle X_i X_j \rangle$ (where $X_i$ stand 
for the mode amplitudes $\Re(\tu^s_\bk)$ and $\Im(\tu^s_\bk)$). The off-diagonal elements
$i \ne j$ are two orders of magnitude smaller than the diagonal elements $i=j$. This indicates that the Fourier amplitudes  $\tu^s_\bk$ are independent variables with Gaussian distribution further verifying the predictions of the previous section. 
}

\subsection{Quasi-equilibrium state and forward flux}

\NEW{In the previous case although it was possible to verify some of the predictions 
of the previous section on $\mP(\bu)$.  However the limited range of wavenumbers did not allow 
to test the predictions on the energy spectrum. To that end we used a series of simulations on a larger grid ($N_G=256$) with $\mI_\mE=1$  varying the viscous coefficient $\nu$ from $\nu=2\cdot 10^{-1}$ to $\nu=10^{-10}$. We must note here that the time to reach saturation from zero initial conditions
is proportional to $T\propto \mE/\mI_\mE$ and can be very large for small values of $\nu$.
For this reason the runs with small $\nu$ started with  random initial conditions 
with energy close to the one predicted.
The same runs were repeated with slightly smaller or large energy to make sure all 
runs converged to the same point.}
\begin{figure}
\centering
\includegraphics[width=0.48\textwidth]{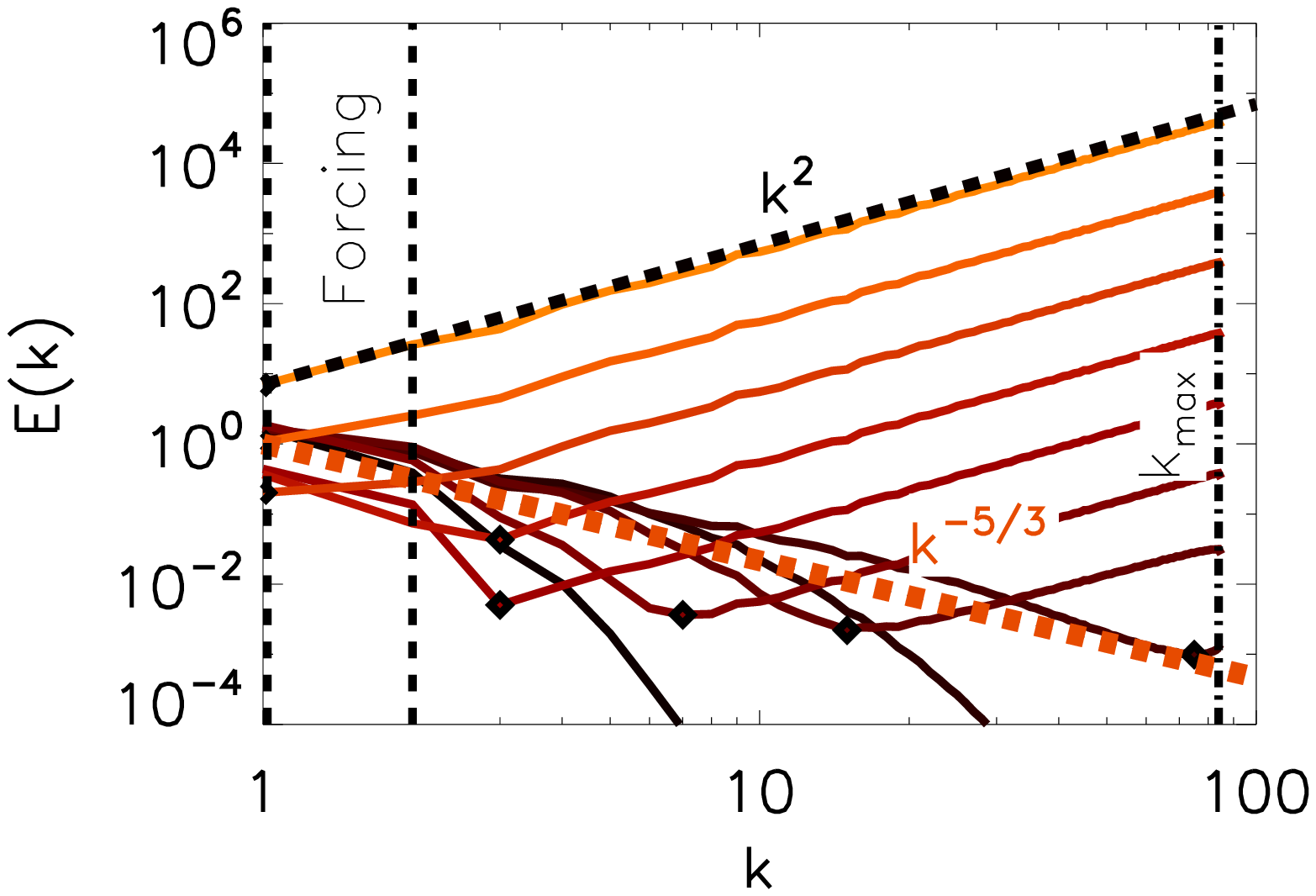}
\includegraphics[width=0.48\textwidth]{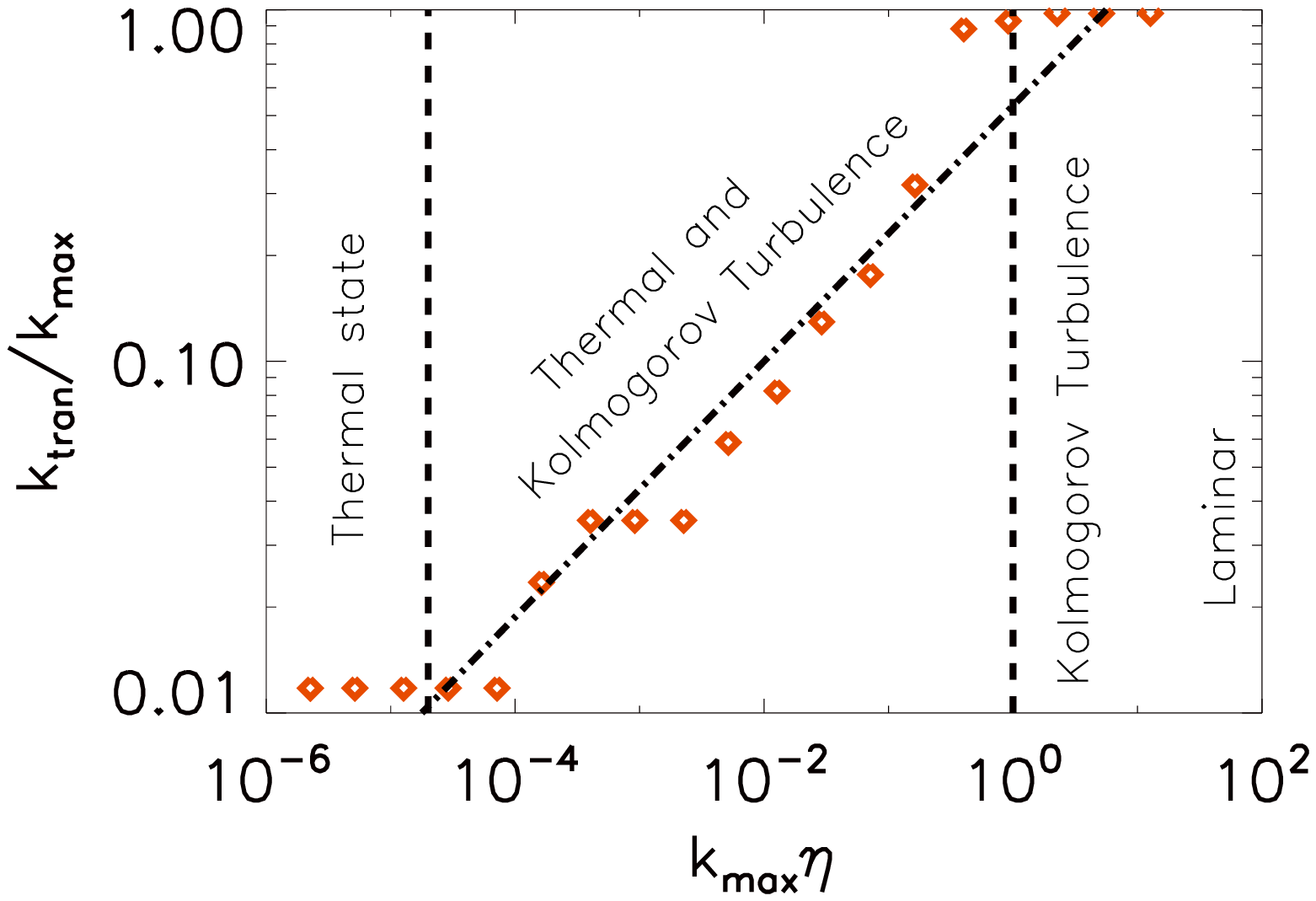}
\caption{Left panel: Energy spectra for six different runs of the TNS equations with $\kmax=85$, $\mI_\mE=1$ and from dark to bright 
$\nu= 10^{-1},\, \nu= 10^{-2}, \,\nu=10^{-3}, \,\nu=10^{-4}, \,\nu=10^{-5}, \,
 \nu= 10^{-6},\, \nu= 10^{-7}, \,\nu=10^{-8}, \,\nu=10^{-9}, \,\nu=10^{-10},$ 
The forcing was restricted in the small wavenumbers and the dissipation in the large wavenumbers as indicated. Right panel: The wavenumber $\kt$ where the Kolmogorov spectrum $k^{-5/3}$ transitions to the thermal spectrum $k^2$ as a function of the Kolmogorov lengthscale $\eta=(\nu^3/\mI_\mE)^{1/4}$. 
Here $\kt$ for the simulations (diamonds) is estimated as the wavenumber that $E(k)$ obtains its minimum. The dashed line gives the prediction in eq. (\ref{ktran}).  }
\label{fig:Spec1}
\end{figure}

The resulting energy spectra are shown in the left panel of figure \ref{fig:Spec1} for ten values of $\nu$. Dark colors indicate large values of $\nu$ while bright values indicate small values of $\nu$. 
\NEW{
For the large values of $\nu$ 
the simulations are `{\it well resolved}'
in the sense that one can observe clearly the dissipation range where there is an exponential decrease of the energy spectrum. 
For $\nu=0.01$ and $\nu=0.001$ (second and third dark line from the bottom) one can also see the formation of an inertial range that displays a negative power-law close to the Kolmogorov prediction $E(k) =C_K \mI_\mE^{2/3} k^{-5/3}$ (where $C_K\simeq 1.6$ is the Kolmogorov constant \cite{sreenivasan1995universality,donzis2010bottleneck}). As the value of $\nu$ is decreased a bottleneck at large wave numbers appears and energy starts to pile  up at the smallest scales of the system. As the value of $\nu$ is decreased further this bottleneck appears to take the form of a positive power-law  close to the thermal equilibrium prediction $E(k)\propto \mA \, k^{2}$. This thermal spectrum occupies more wavenumbers as the viscosity is decreased until all wavenumbers follow  this scaling.
At the smallest value of $\nu$ the result is compared with the asymptotic result obtained in the previous section. }
\NEW{The proportionality coefficient $\mA$ can be estimated from eq. \ref{fspec} to be $\mA = 5 \mI_\mE / (2\nu \kmax^5)$ that guaranties that the energy balance condition $2\nu \int_0^{\kmax} k^2 E(k)dk = \mI_\mE$ is satisfied.} 
Matching the thermal with the Kolmogorov spectrum we obtain that the transition occurs at the wavenumber
\beq 
\kt = \kmax \left( \frac{2C_K}{5}\right)^{3/11}  (\kmax \eta)^{4/11} \label{ktran}
\eeq
where $\eta=(\nu^3/\mI_\mE)^{1/4}$ is the Kolmogorov length-scale.
The right panel of fig. \ref{fig:Spec1} shows a comparison of this estimate with $\kt$ measured from the spectra as the wavenumber at which $E(k)$ obtains its minimum.
The scaling agrees very well with the results from the simulations.

\begin{figure}
\centering
\includegraphics[width=0.88\textwidth]{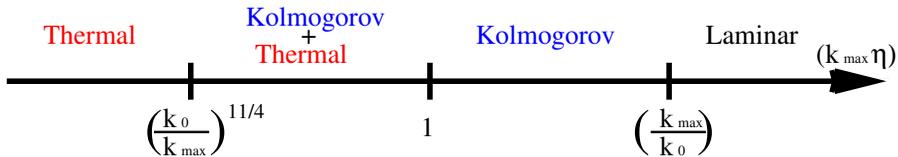}
\caption{ Turbulence behavior as a function of $\kmax \eta$.}
\label{fig:thermal_line}
\end{figure}

\NEW{
In summary for values of $\kmax\eta$ larger than $k_0/\kmax$ ({\it ie} $k_0\eta\gg1$) the flow is laminar displaying an exponential spectrum. 
For $1 \ll \kmax\eta \ll k_0/\kmax $
there is the formation of an inertial range 
where the Kolmogorov spectrum dominates followed by the dissipation range.
If $(k_0/\kmax)^{11/4} \ll \kmax\eta \ll 1 $
(so that $\kt \gg k_0$) then the dissipative range no longer exist and there is coexistence of the Kolmogorov spectrum followed by a thermalized spectrum at large wavenumbers.
Finally for $\kmax\eta \ll (k_0/\kmax)^{11/4}$
the system is in the quasi-equilibrium thermal state predicted in the previous section. These result are summarized in the sketch in fig. \ref{fig:thermal_line}.}

The transition from a Kolmogorov spectrum to a thermal one resembles a lot the time evolution of the TEE studied in \cite{cichowlas2005effective}. For the TEE at early times a $k^{-5/3}$ energy spectrum develops as energy is transferred to larger and larger wavenumbers. When the maximum wavenumber $k_{max}$ is reached, the thermalized energy spectrum starts to develop 
displaying at intermediate times both spectral slopes $k^{-5/3}$ and $k^{2}$. 
The difference with the present runs is that in the Euler case the transition occurs as time is increased  while in the present case we only consider the steady state, and vary the value of $\kmax\eta$. 

Similarities can also be found with the recent work on a time-reversible version of the Navier-Stokes \citep{gallavotti1996equivalence}. Using shell models \cite{biferale2018equivalence} and three dimensional simulations \cite{shukla2018phase}.
of the time-reversible Navier-Stokes these authors found
similar transitions from a Kolmogorov to a thermal quasi-equilibrium with the formation of both spectra depending on the parameter regime. The results were interpreted in terms of a phase transition, a possibility that could further explored for the present work as well.

\begin{figure}
\centering
\includegraphics[width=0.47\textwidth]{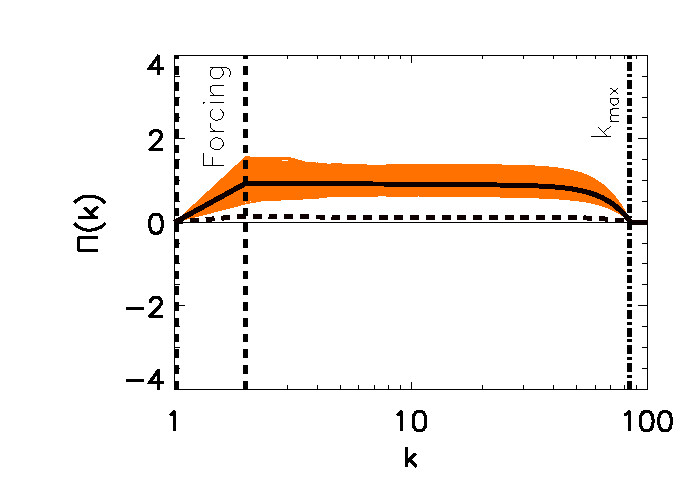}
\includegraphics[width=0.47\textwidth]{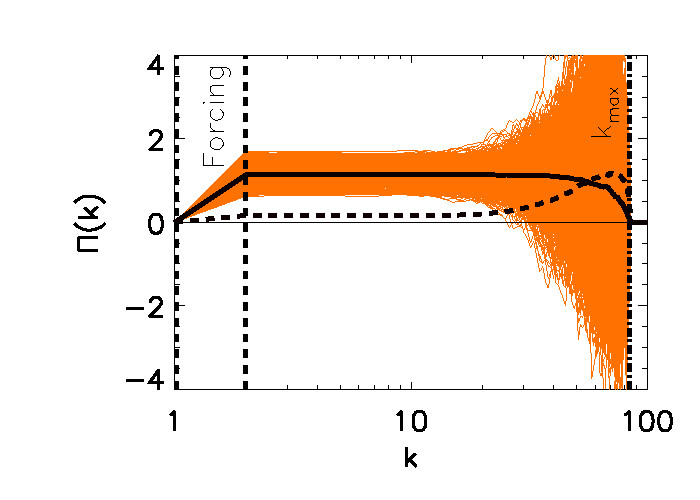}
\includegraphics[width=0.47\textwidth]{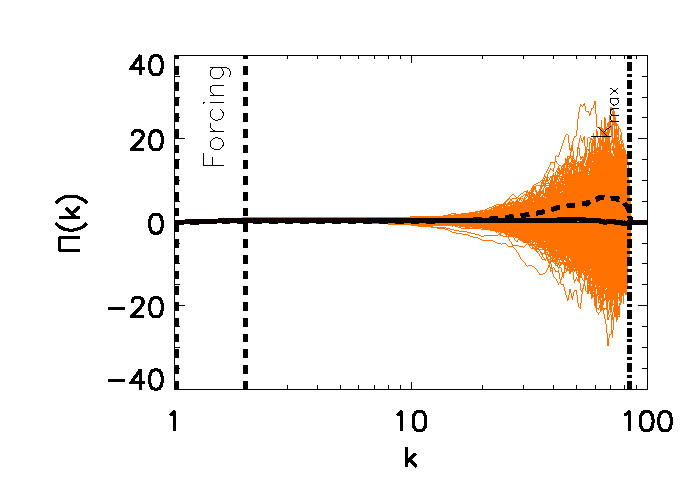}
\includegraphics[width=0.47\textwidth]{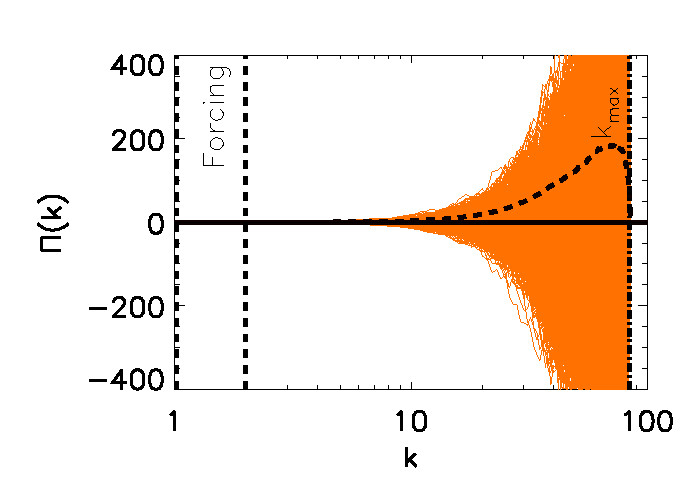}
\caption{ Time average energy fluxes (dark line) and instantaneous energy fluxes (bright lines) for three different values of the viscosity 
          $\nu=10^{-3}$ (top left panel), 
          $\nu=10^{-5}$ (top right panel),
          $\nu=3 \cdot 10^{-5}$ (bottom left panel),
          $\nu=3 \cdot 10^{-6}$ (bottom right panel).
The dashed line shows the variation amplitude of the flux. (Note the change in the scale of the y-axis in the bottom figures)}
\label{fig:Flux1}
\end{figure}

\begin{figure}
\centering
\includegraphics[width=0.47\textwidth]{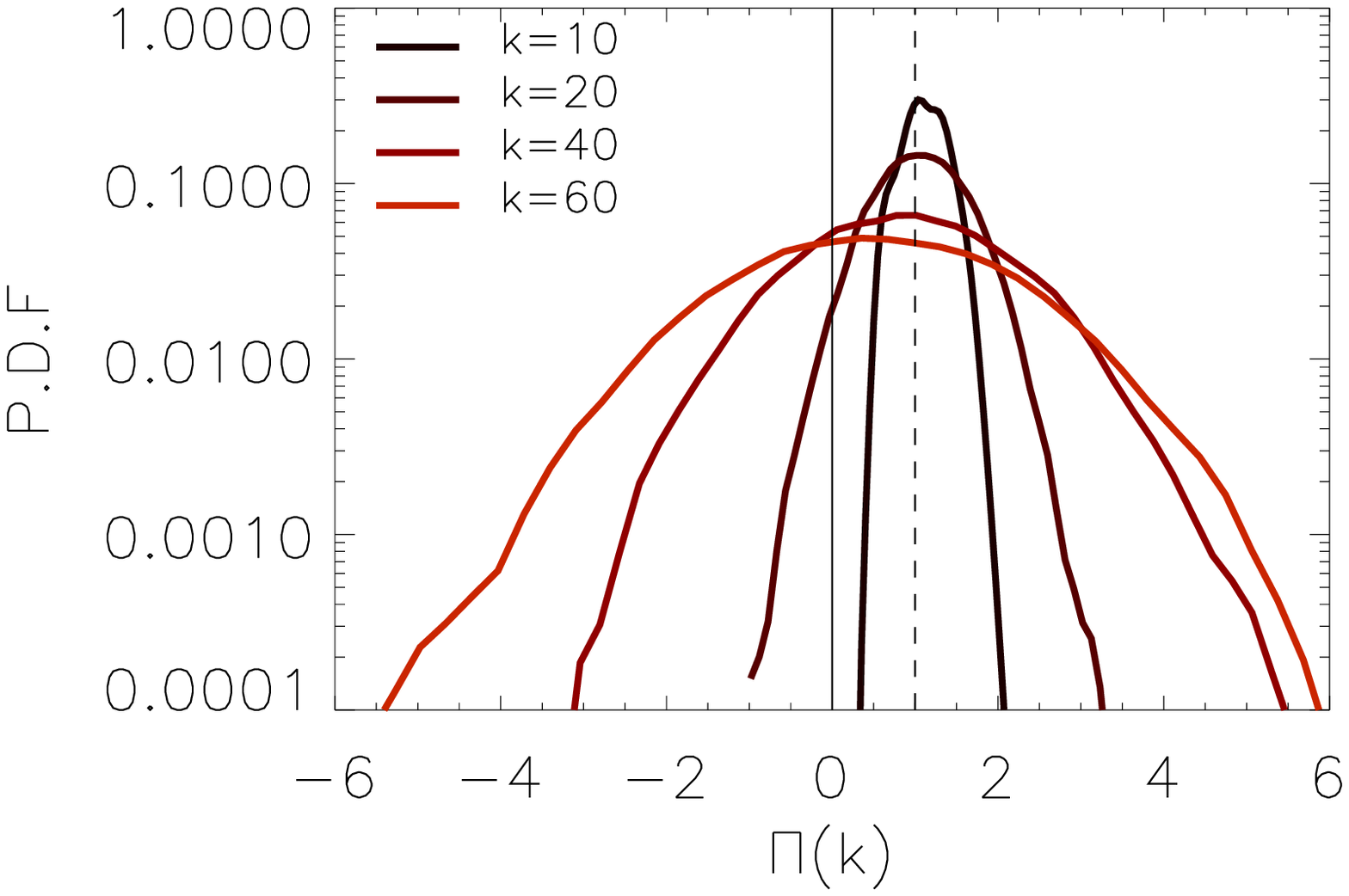}
\includegraphics[width=0.47\textwidth]{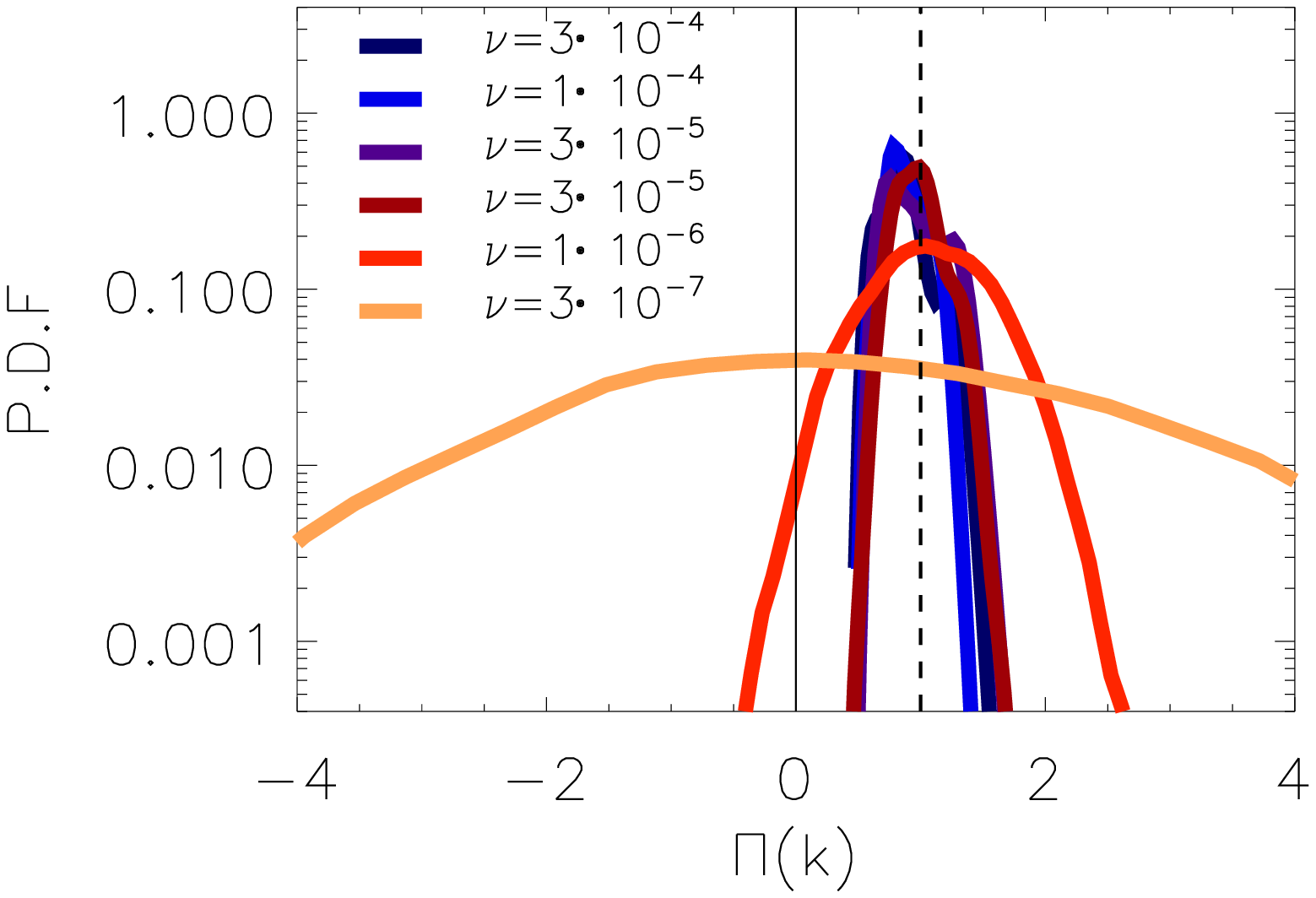}
\caption{ Left panel: probability distribution function of the energy flux
          for $\nu=10^{-5}$ at four different wavenumbers.
          Right panel: probability distribution function of the energy flux
          at $k=20$ and different values of viscosity.
The dashed line shows the variation amplitude of the flux.}
\label{fig:PFlux1}
\end{figure}

We note that the time averaged flux from large scales to small scales is constant in the inertial range. What varies as we change the value of $\kmax\eta$ is the amplitude of the fluctuations of the flux around this mean value of the flux. This is displayed in figure \ref{fig:Flux1} where the mean flux is shown with dark line together with the instantaneous fluxes at different times with bright colors for four different values of $\nu$. 
For $\nu$ in the Kolmogorov turbulence regime the fluctuations of the flux are concentrated around the mean value without large deviations from it. As the value of the viscosity is \NEW{decreased} the fluctuations at large wavenumbers are increased until finally the fluctuations of the mean flux are orders of magnitude larger than the averaged flux for all wave numbers. 
\NEW{The pdfs of the energy flux for different wavenumbers and values of viscosity are shown in fig. \ref{fig:PFlux1}.
Note that for a flow in equilibrium all third order quantities like the flux have zero value. 
In this case the mean value of the flux remains fixed but the variance of the fluctuations as $\nu$ is decreased (right panel of fig. \ref{fig:PFlux1}) or $k$ is increased (left panel of fig. \ref{fig:PFlux1}) increase. Thus compared to the variance of the fluctuations
the mean flux becomes negligible and thus the flow is in {\it quasi-equilibrium}.}

\subsection{Quasi-equilibrium state and inverse flux}

\begin{figure}
\centering
\includegraphics[width=0.48\textwidth]{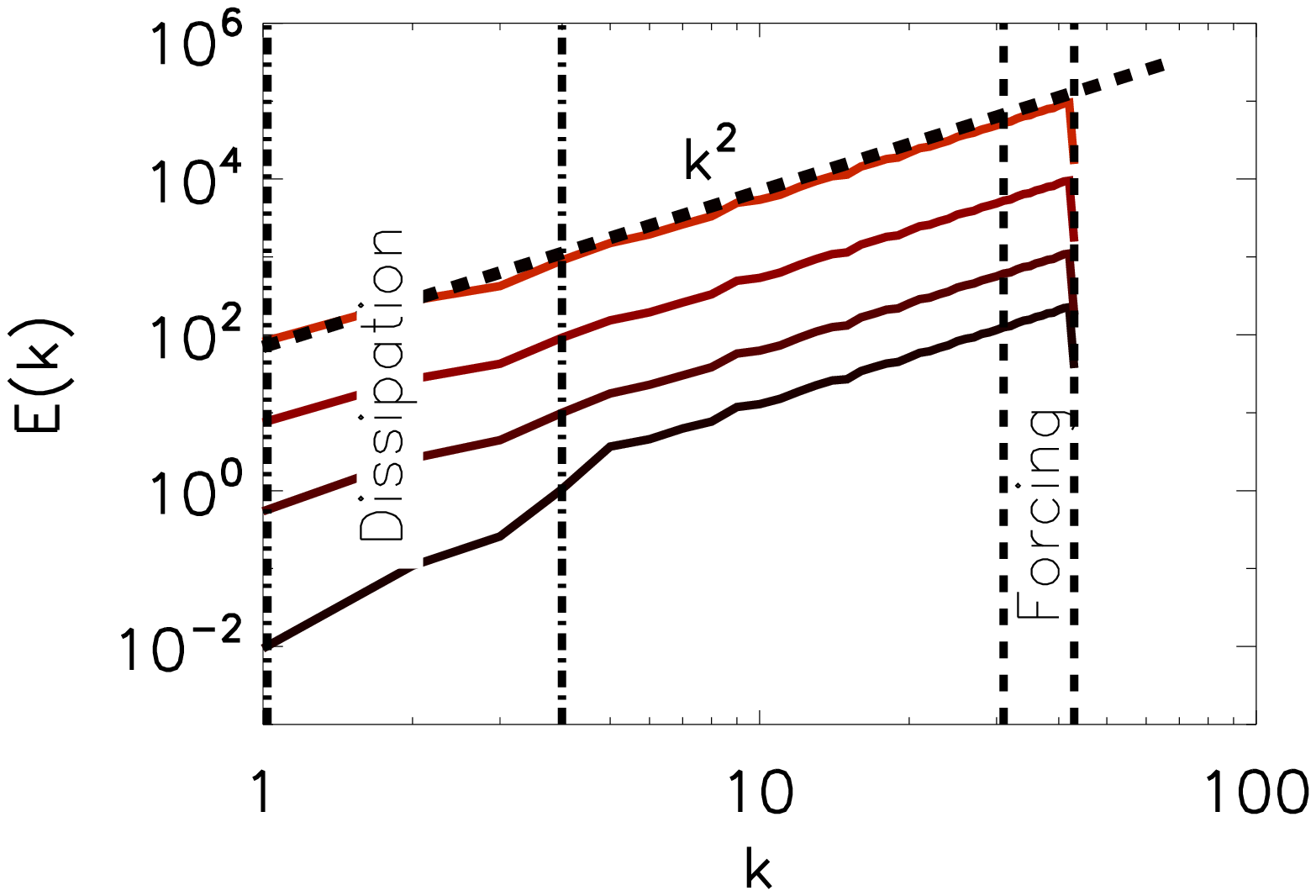}
\includegraphics[width=0.48\textwidth]{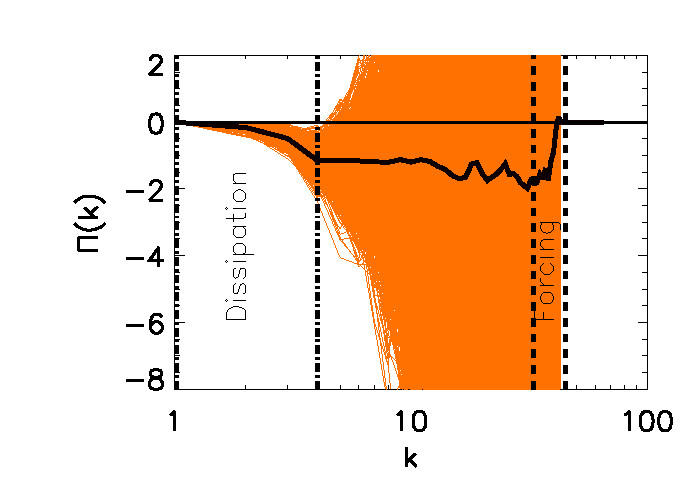}
\caption{ Left panel: Energy spectra from simulation forced at small scales and dissipated at large scales
          as indicated. For all runs $\mI_\mE=1$ and from dark to bright $\nu=10,\, \nu=1, \,\nu=0.1,\, \nu=0.01$.
          Right panel: Time average energy flux (dark line) and instantaneous energy fluxes (bright lines) for
          $\nu=1$.}
\label{fig:SpecFlux2}
\end{figure}

\NEW{
In most three dimensional turbulent flows the scales larger than the forcing scale 
are close to an equilibrium state \citep{alexakis2018thermal}. In many instances however there is a change 
of dynamics at large enough scales and the flow is constrained to two dimensional dynamics 
(like for example  turbulence in thin layers or in rotating flows) 
where energy tends to cascade inversely \citep{alexakis2018cascades}. 
There is thus a transition from a thermal state with zero energy flux, to a state 
that has a finite inverse flux of energy. Close to such transition (unless the transition is discontinuous)
the system has to be close to the equilibrium state with an inverse energy flux.
With this motivation we examine the case where the forcing is located at small scales while the dissipation
is limited in the large scales so that there is an net inverse transfer of energy.} 

\NEW{
We have thus performed simulations on a $N_G=128$ numerical grid that leads to a $\kmax=48$ forcing at wavenumbers in the range $(k_{_F}=31,k_{_F}'=45)$ with unit energy injection rate.
At difference with the previous simulations the dissipation limited only to small wavenumbers that satisfy $1\le |\bk| \le k'=4$.
The energy is thus injected at small scales and dissipated at large scales forcing that way an inverse transfer of energy.}

\begin{figure}
\centering
\includegraphics[width=0.48\textwidth]{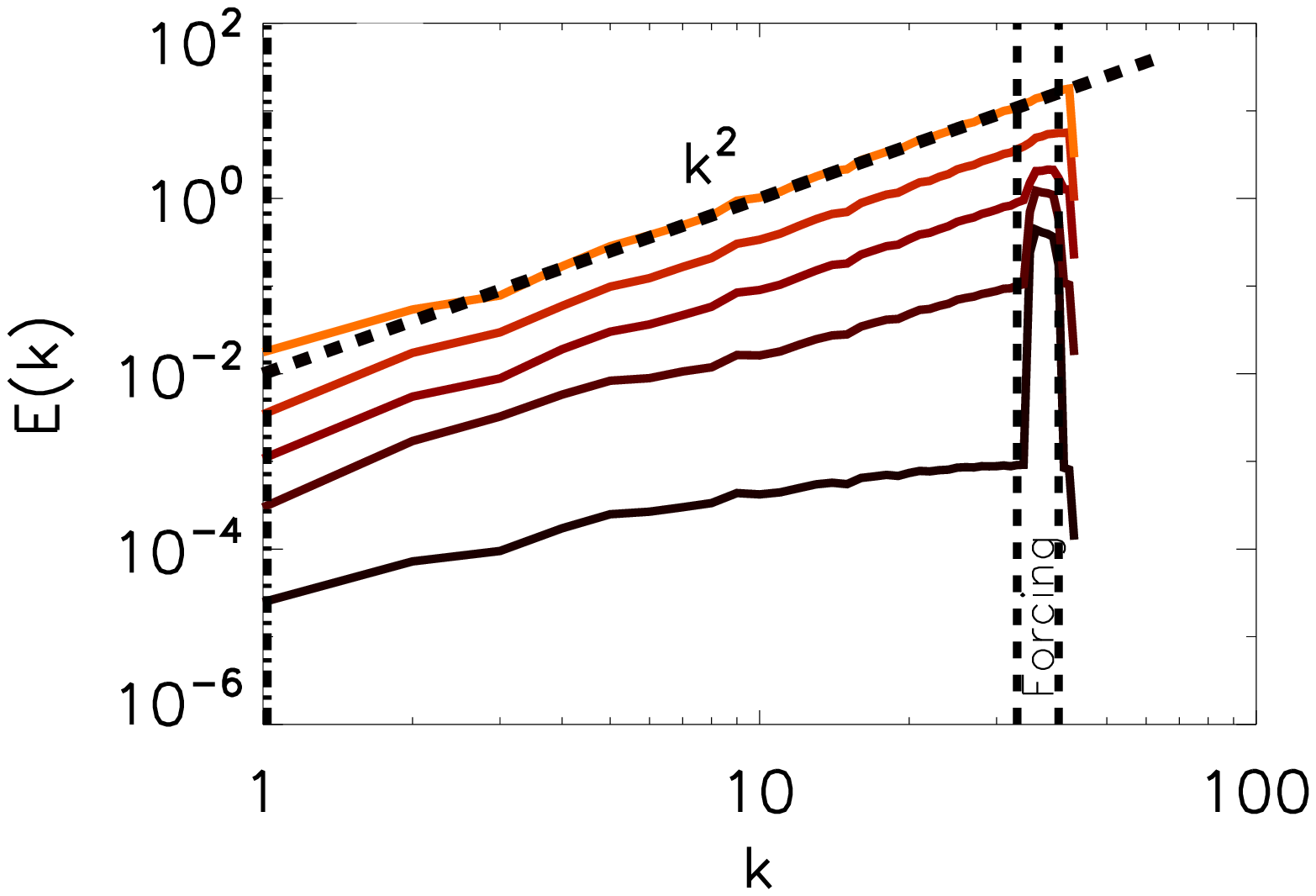}
\includegraphics[width=0.48\textwidth]{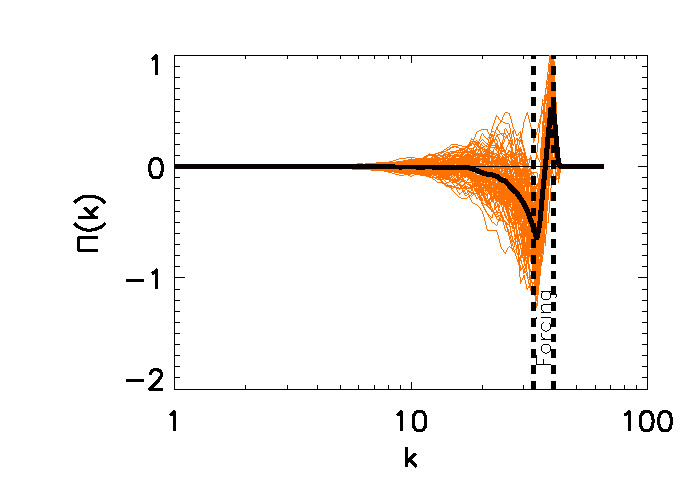}
\caption{ Left panel: Energy spectra from simulation forced at small scales and dissipated by regular viscosity. For all runs $\mI_\mE=1$ and from dark to bright $\nu=0.01,\, \nu=0.003, \,\nu=0.001,\, \nu=0.0003, \,\nu=0.00001$.
          Right panel: Time average energy flux (dark line) and instantaneous energy fluxes (bright lines) for
          $\nu=0.00001$.}
\label{fig:SpecFlux3}
\end{figure}

The spectra for four different values of $\nu$ are shown in the left panel of fig. \ref{fig:SpecFlux2}. 
The forcing and dissipation shells are indicated in the graph.
In this case a spectrum  $E(k)=\mA k^2$ forms for all values of $\nu$
with small changes at the dissipation wave numbers for large values of $\nu$. 
For small values o $\nu$ the amplitude of the spectrum $\mA$ is such so that the energy balance is satisfied
that leads to $\mA \simeq  5 \mI_\mE/(2\nu k'^5)$ where $k'$ is the maximum wavenumber where the dissipation acts.
The dashed line shows this prediction for the smallest value of $\nu$.
The flux fluctuations shown in the right panel of the same figure (fig. \ref{fig:SpecFlux2}) are always dominant.  
We note however that their averaged value leads to a constant negative flux in the $k^2$ 
inertial range where no forcing or dissipation are present.
\NEW{ It is worth noting that in these quasi-equilibrium states the direction of the transfer of
energy is not determined by the nonlinear term as in Kolmogorov turbulence but only  
by the location of the energy source and sink. If regular viscosity is used, 
although a thermalized state is still reached as predicted by the previous section,
there is no net inverse flux of energy. This is demonstrated in fig. \ref{fig:SpecFlux3}
where the energy spectra and the energy flux is shown from simulations with regular viscosity forced at small scales.
}

\section{Conclusions}                            

\NEW{In this work we examined the spectrally truncated Navier-Stokes equations flows that are close to equilibrium.}
We showed analytically that in the limit of small viscosity the statistically steady state of these flows converge to the \cite{kraichnan1973helical} solutions.
We note that our prediction is not just for the spectrum in eq. (\ref{fspec}) but for the full probability distribution $\mP(\bu)$ for the system to find itself in a state $\bu$ given in eq. (\ref{final}).  

\NEW{The derivation was based on two major assumptions.}
First we assumed ergodicity for the solutions of the TEE. This assumption appears in most calculations of classical statistical physics that although it can only be proved in very few systems it appears to be  a plausible one for many systems with large numbers of degrees of freedom. For the TEE it appears at least to be in agreement with the results of numerical simulations. The second assumption we made was to neglect the effect of the second invariant: the helicity. This assumption was made in order to simplify the (rather involved) calculation. Had we kept the effect of helicity
then the zeroth order solution $\mP_0(\bu)$ would have been reduced to a function of two variables $f(\mE,\mH)$ and we would have ended up with an elliptic partial differential equation to solve for $f$.  However the presence of helicity would break the spherical symmetry in phase space that allowed us to calculate the involved integrals. The calculation is still feasible but much more lengthier and we leave it for future work. 

\NEW{
The numerical investigation verified our analytical results and demonstrated that the Fourier 
amplitudes $\tu_\bk^s$ indeed become independent Gaussian variables,
and that the energy distribution and energy spectrum approach that of the analytic predictions 
as the asymptotic limit is reached.
Furthermore, the numerical investigation also shed light on how the TNS system transitions from the classical Kolmogorov turbulence state to the thermalized solutions of \cite{kraichnan1973helical}  as the $\kmax\eta$ is varied and led to precise predictions on when the transition takes place. The transitions are compactly summarized in fig. \ref{fig:thermal_line} where the three different states 
``{\it laminar}", ``{\it turbulent}" and ``{\it thermal}" as well as the transitions from one state to the other are clearly marked.
Finally it was also demonstrated that these quasi equilibrium states are present when  
the dissipation is localized in the small wave numbers and the forcing at large wavenumbers forming an inverse flux of energy.}

\NEW{
It is worth noting that a fixed amplitude flux (positive or negative) was always present in our simulations and that it was determined by the injection rate. However, as the quasi-equilibrium state is approached,  the amplitude of the velocity fluctuations increases. 
These velocity fluctuations then lead to fluctuations of the energy flux to have variance 
that is much larger then the mean value making this mean flux a sub-dominant. 
Furthermore, the mean energy in the quasi-equilibrium state scales like $\langle \mE \rangle \propto 1/\nu $ (see eq. \ref{eq:meanE})
and therefore in the small viscosity limit the energy dissipation normalized by $u_{rms}^3k_0$ 
\[   \lim_{\nu \to 0}   \frac{\langle \mI_\mE \rangle}{ \langle 2 \mE \rangle^{2/3} k_0} =0
\]
becomes zero. Thus, opposed to Kolmogorov turbulence, there is no finite (normalized) energy dissipation rate in the zero viscosity limit.
We have thus to distinguish the processes of energy transfer in these quasi-equilibrium states (sometimes referred as ``{\it warm cascades}") from the energy cascade that is met classical turbulent flows. These states are dominated by fluctuations and the direction of the energy transfer is not determined by the properties of the non-linearity
but only the location of the sources and sinks of energy and in general is a non-local process (see fig. 14 in \cite{alexakis2018thermal}).
}

There are many directions in which the present results can be pursued further. First of all including the effect of helicity is crucial to have a complete description of the system. Moreover carrying out the calculation at the next order so that statistics of the fluxes can also be calculated would be equally desirable. Finally, extending these results to two-dimensional flows, where the equilibrium states can take the form of large scale condensates, is another possible direction.  Such calculations, although considerably longer than the ones presented here, should still be feasible and we hope to address them in our future work.

\acknowledgments 
This work was granted access to the HPC resources of MesoPSL financed by the Region 
Ile de France and the project Equip@Meso (reference ANR-10-EQPX-29-01) of the programme Investissements d'Avenir supervised by the Agence Nationale pour la Recherche and the HPC resources of GENCI-TGCC \& GENCI-CINES (Project No. A0050506421) where the present numerical simulations have been performed. This work has also been supported by the Agence nationale de la recherche (ANR DYSTURB project No. ANR-17-CE30-0004). This work was also supported by the research Grant No. 6104-1 from Indo-French Centre for the Promotion of Advanced Research (IFCPAR/CEFIPRA).

\appendix 

\bibliographystyle{jfm}
\bibliography{thermal}



\end{document}